\input harvmac

%
\def\abstract#1{
\vskip .5in\vfil\centerline
{\bf Abstract}\penalty1000
{{\smallskip\ifx\answ\bigans\leftskip 1pc \rightskip 1pc 
\else\leftskip 1pc \rightskip 1pc\fi
\noindent \abstractfont  \baselineskip=12pt
{#1} \smallskip}}
\penalty-1000}
\baselineskip=15pt plus 2pt minus 1pt

\def\hth/#1#2#3#4#5#6#7{{\tt hep-th/#1#2#3#4#5#6#7}}
\def\nup#1({Nucl.\ Phys.\ $\us {B#1}$\ (}
\def\plt#1({Phys.\ Lett.\ $\us  {B#1}$\ (}
\def\cmp#1({Comm.\ Math.\ Phys.\ $\us  {#1}$\ (}
\def\prp#1({Phys.\ Rep.\ $\us  {#1}$\ (}
\def\prl#1({Phys.\ Rev.\ Lett.\ $\us  {#1}$\ (}
\def\prv#1({Phys.\ Rev.\ $\us  {#1}$\ (}
\def\mpl#1({Mod.\ Phys.\ Let.\ $\us  {A#1}$\ (}
\def\atmp#1({Adv.\ Theor.\ Math.\ Phys.\ $\us  {#1}$\ (}
\def\ijmp#1({Int.\ J.\ Mod.\ Phys.\ $\us{A#1}$\ (}
\def\jhep#1({JHEP\ $\us {#1}$\ (}

\def\bb#1{{\bar{#1}}}

\def\cx#1{{\cal #1}}

\def\hx#1{{\hat{#1}}}
\def\vx#1{\vec{#1}}
\def\rmx#1{{\rm #1}}
\def\us#1{\underline{#1}}
\def\fc#1#2{{#1\over #2}}
\def\frac#1#2{{#1\over #2}}

\def\br{\hfill\break}
\def\noi{\noindent}

\def\al{\alpha}\def\om{\omega}
\def\p{\partial}

\def\CY{Calabi--Yau }

\def\Gah{{\hx\Gamma}}
\def\zh{{\hx z}}

\def\bra#1{|#1\rangle_{RR}}

\def\Om{\Omega}\def\Si{\Sigma}
\def\p{\partial}
\def\CC{{\bf C}}\def\ZZ{{\bf Z}}
\def\CY{Calabi--Yau }

\def\bb#1{\bar{#1}}

\def\phix#1{\phi^{(#1)}}

\def\Ga{\Gamma}\def\Si{\Sigma}
\def\MM{\cx M _{_{\!\cx N=1}}}\def\Mm{\cx M _{_{\!\cx N=2}}}

\def\Roc{\cx R_{oc}}

\def\doubref#1#2{\refs{{#1},{#2}}}

\def\dd{\cdot}


\def\CC{{\bf C}}

\def\xt(#1){\theta_{#1}}
\def\xa(#1){n_{#1}}
\def\xz(#1){z_{#1}}

\def\vx#1{\vec{#1}}

\ifx\answ\bigans\else\fi
%
\def\Z{Y}\def\z{z}
\def\Y{B}
\def\C{{B}}

\def\Phiv#1{{\vx \Phi}^{(#1)}}
\def\weqn#1{\xdef #1{(\noexpand\hyperref{}%
{equation}{\secsym\the\meqno}%
{\secsym\the\meqno})}\eqno(
{\secsym\the\meqno}{\secsym\the\meqno})\eqlabeL#1%
\writedef{#1\leftbracket#1}\global\advance\meqno by1}
\def\weqnalign#1{\xdef #1{\noexpand\hyperref{}{equation}%
{\secsym\the\meqno}{(\secsym\the\meqno)}}%
\writedef{#1\leftbracket#1}%
{\secsym\the\meqno}{\e@tf@ur#1}\eqlabeL{#1}%
\global\advance\meqno by1}
\def\abstract#1{
\vskip .5in\vfil\centerline
{\bf Abstract}\penalty1000
{{\smallskip\ifx\answ\bigans\leftskip 1pc \rightskip 1pc
\else\leftskip 1pc \rightskip 1pc\fi
\noindent \abstractfont  \baselineskip=12pt
{#1} \smallskip}}
\penalty-1000}
\def\eprt#1{{#1}}
\def\nihil#1{{\sl #1}}
\lref\MiBo{See e.g.:
\nihil{Essays on mirror manifolds}, (S.\ Yau, ed.),
International Press 1992;
\nihil{Mirror symmetry II}, (B.\ Greene et al, eds.),
International Press 1997.}
\lref\Kar{M. Karoubi and C. Leruste, \nihil{Algebraic topology
via differential geometry}, Cambridge Univ. Press,
Cambridge, 1987.}
\lref\auto{%
S.~Ferrara, D.~L\"ust, A.~D.~Shapere and S.~Theisen,
\nihil{Modular Invariance In Supersymmetric Field Theories,}
Phys.\ Lett.\ B {\bf 225}, 363 (1989).}

\lref\sugra{
E.~Cremmer, C.~Kounnas, A.~Van Proeyen, J.~P.~Derendinger, S.~Ferrara, B.~de Wit and L.~Girardello,
\nihil{Vector Multiplets Coupled To N=2 Supergravity: 
Superhiggs Effect, Flat Potentials And Geometric Structure,}
Nucl.\ Phys.\ B {\bf 250}, 385 (1985);\br
B.~de Wit, P.~G.~Lauwers and A.~Van Proeyen,
\nihil{Lagrangians Of N=2 Supergravity - Matter Systems,}
Nucl.\ Phys.\ B {\bf 255}, 569 (1985).
}

\lref\sgCY{R.L. Bryant and P.A. Griffiths, in {\it Arithmetic and
Geometry}, eds. M. Artin and J. Tate, Birkh\"auser, Boston, 1983;\br
A.~Strominger,
\nihil{Special Geometry,}
Commun.\ Math.\ Phys.\  {\bf 133}, 163 (1990);\br
P.~Candelas and X.~de la Ossa,
\nihil{Moduli Space Of Calabi-Yau Manifolds,}
Nucl.\ Phys.\ B {\bf 355}, 455 (1991).
}

\lref\SSTwo{K.~Becker, M.~Becker and A.~Strominger,
\nihil{Five-branes, membranes and nonperturbative string theory,}
Nucl.\ Phys.\ B {\bf 456}, 130 (1995), hep-th/9507158.}
\lref\copa{W. Lerche, P.Mayr and N. Warner, 
\nihil{$N=1$ Special Geometry, Mixed Hodge Variations and Toric Geometry}, 
to appear.}
\lref\AVi{
{M.\ Aganagic and C.\ Vafa, 
\nihil{Mirror symmetry, D-branes and counting holomorphic discs,}
\eprt{hep-th/0012041}. 
}}

\lref\AVii{M.\ Aganagic, A.\ Klemm and C.\ Vafa, 
\nihil{Disk instantons, mirror symmetry and the duality web,}
 Z.\ Naturforsch.\ A{\bf 57} 1 (2002), 
\eprt{hep-th/0105045}. 
}

\lref\BCOV{M.\ Bershadsky, S.\ Cecotti, H.\ 
Ooguri and C.\ Vafa, 
\nihil{Kodaira-Spencer theory of gravity and exact 
results for quantum string amplitudes,}
 Commun.\ Math.\ Phys.\ {\bf 165} 311 (1994), 
\eprt{hep-th/9309140}. 
}

\lref\CeVa{S.\ Cecotti and C.\ Vafa, 
\nihil{Topological antitopological fusion,}
 Nucl.\ Phys.\ B{\bf 367} 359 (1991). 
}

\lref\Del{P. Deligne, \nihil{Theori\'e de Hodge I-III},
Actes de Congr\`es international de Mathematiciens (Nice,1970),
Gauthier-Villars {\bf 1} 425 (1971);
Publ. Math. IHES {\bf 40} 5 (1971);
Publ. Math. IHES {\bf 44} 5 (1974).}

\lref\Grifdeq{P.A. Griffiths, 
\nihil{A theorem concerning the differential equations satisfied
by normal functions associated to algebraic cycles},
Am. J. Math {\bf 101} 96 (1979).}

\lref\Guk{S.~Gukov,
\nihil{Solitons, superpotentials and calibrations,}
Nucl.\ Phys.\ B {\bf 574}, 169 (2000),
hep-th/9911011.}
\lref\GVW{
S.\ Gukov, C.\ Vafa and E.\ Witten, 
\nihil{CFT's from Calabi-Yau four-folds,}
 Nucl.\ Phys.\ B{\bf 584} 69 (2000), 
Erratum-ibid.\ B{\bf 608} 477 (2001),
\eprt{hep-th/9906070}.}

\lref\KKLM{S.\ Kachru, S.\ Katz, A.\ E.\ Lawrence and J.\ McGreevy, 
\nihil{Open string instantons and superpotentials,}
 Phys.\ Rev.\ D{\bf 62} 026001 (2000), 
\eprt{hep-th/9912151}. 
}

\lref\KLMVW{A.\ Klemm, W.\ Lerche, P.\ Mayr, C.\ Vafa and N.\ P.\ Warner, 
\nihil{Self-Dual Strings and N=2 Supersymmetric Field Theory,}
 Nucl.\ Phys.\ B{\bf 477} 746 (1996), 
\eprt{hep-th/9604034}. 
}

\lref\LM{W.\ Lerche and P.\ Mayr, 
\nihil{On $N=1$ mirror symmetry for open type II strings,}
\eprt{hep-th/0111113}. }

\lref\LVW{
W.~Lerche, C.~Vafa and N.~P.~Warner,
\nihil{Chiral Rings In N=2 Superconformal Theories,}
Nucl.\ Phys.\ B {\bf 324}, 427 (1989).
}

\lref\PM{P.\ Mayr, 
\nihil{$N=1$ mirror symmetry and open/closed string duality,}
Adv.\ Theor.\ Math.\ Phys.\  {\bf 5}, 213 (2001),
\eprt{hep-th/0108229}.}

\lref\PMsusy{P.\ Mayr, 
\nihil{On supersymmetry breaking in string theory and 
its realization in brane worlds,}
 Nucl.\ Phys.\ B{\bf 593} 99 (2001), 
\eprt{hep-th/0003198}. 
}

\lref\TV{T.\ R.\ Taylor and C.\ Vafa, 
\nihil{RR flux on Calabi-Yau and partial supersymmetry breaking,}
 Phys.\ Lett.\ B{\bf 474} 130 (2000), 
\eprt{hep-th/9912152}. 
}

\lref\OV{H.~Ooguri and C.~Vafa,
\nihil{Knot invariants and topological strings,}
Nucl.\ Phys.\ B {\bf 577}, 419 (2000), hep-th/9912123.}

\lref\CVlargeN{C.~Vafa,
\nihil{Superstrings and topological strings at large N,}
J.\ Math.\ Phys.\  {\bf 42}, 2798 (2001), hep-th/0008142.}

\lref\WitCS{E.\ Witten, 
\nihil{Chern-Simons gauge theory as a string theory,}
\eprt{hep-th/9207094}.} 

\lref\WQCD{E.\ Witten, 
\nihil{Branes and the dynamics of QCD},
 Nucl.\ Phys.\ B{\bf 507} 658 (1997), 
\eprt{hep-th/9706109}. 
}

\vskip-2cm
\Title{\vbox{
\rightline{\vbox{\baselineskip12pt\hbox{CERN-TH/2002-174}
         \hbox{hep-th/0207259}}}\vskip0cm}}
{Holomorphic $N=1$ Special Geometry}
\vskip -0.99cm
\centerline{\titlefont of Open--Closed Type II Strings}

\abstractfont 
\vskip 0.8cm
\centerline{W. Lerche$^\flat$, P. Mayr$^\flat$ and N. Warner$^\sharp$}

\vskip 1.5cm
\centerline{$^\flat$ \ninepoint
CERN Theory Division, CH-1211 Geneva 23, Switzerland}
\centerline{$^\sharp$ \ninepoint Department of Physics and Astronomy,
University of Southern California, }
\centerline{\ninepoint Los Angeles, CA 90089-0484, USA}
\vskip 0.3cm
\abstract{
We outline a general
geometric structure that underlies the $\cx N=1$ superpotentials of 
a certain class of flux and brane configurations in type II string compactifications
on Calabi-Yau threefolds. This ``holomorphic $\cx N=1$ special geometry'' is in
many respects comparable to, and in a sense an extension of,
the familiar special geometry in $\cx N=2$ supersymmetric type II string
compactifications. It puts the computation of the instanton-corrected 
superpotential $W$ of the four-dimensional $\cx N=1$ string effective
action on a very similar footing as the familiar
computation of the $\cx N=2$ prepotential $\cx F$ via mirror symmetry.
In this note we present some of the main ideas and
results, while more details as well as some explicit computations will appear
in a companion paper.
}
\vskip1cm
\Date{\vbox{\hbox{ {July 2002}}
}}
\goodbreak

\newsec{Introduction}
As is well-known,
the manifold $\Mm$ of scalar vev's $z_a$ of the vector multiplets of a
four-dimensional $\cx N=2$ supergravity is characterized by a
``special geometry''. In the context of an effective 
supergravity theory obtained from the compactification of a type II 
string on a \CY manifold $X$, the $\cx N=2$ special geometry can be
understood from (at least) three different points of view:
\item{i)}
as a consequence of local $\cx N=2$ space-time supersymmetry \sugra;
\item{ii)}
from the structure of the underlying 2d 
topological field theory (TFT) on the string world-sheet, 
whose correlation functions are
summarized by the $\cx N=2$ effective supergravity \BCOV; 
\item{iii)} 
for type IIB strings, in terms of the Hodge structure
on the middle cohomology of $X$ which varies with 
the complex structure moduli $z_a$ \sgCY.

\noi The last formulation is particularly important, as it allows
to determine the {\it exact} non-perturbative effective action
(up to two derivatives) for the vector multiplets from classical geometric data; that is,
 from the periods of the holomorphic (3,0) form, integrated over an 
integral basis for the middle homology
$H_3(X,\ZZ)$. This fact has been
exploited with large success in the context of closed string 
mirror symmetry \MiBo.

The purpose of this letter is to report similar
results on a special geometry of the holomorphic $F$-terms of 
certain $\cx N=1$ supersymmetric string compactifications. 
Specifically, these are type II compactifications on \CY
manifolds with extra fluxes and (D-)branes. As is evident, this
holomorphic $\cx N=1$ special geometry will not be
a general consequence of $\cx N=1$ space-time supersymmetry, 
but really a property of the string effective theory.
However,  as will be discussed, 
the counterparts of the above items $ii)$ and $iii)$ still
exist in this phenomenologically relevant class of superstring
theories. Some aspects of the special geometry have been
studied already in \doubref\PM\LM, based on 
a class of D-brane geometries defined and studied in the 
important work \AVi.

In the following, we describe the $\cx N=1$ special geometry
as the consequence of systematically incorporating fluxes
and branes into the familiar ideas and methods of 
mirror symmetry for \CY 3-folds. 
Starting with the $\cx N=1$ counterpart of item ii) above,
the TFT description involves the chiral ring $\Roc$ of the open-closed B-model,
and an integrable, topological connection on the space of 2d 
RR ground states generated by the elements of $\Roc$. These concepts are,
as in the case of $\cx N=2$ special geometry, particular aspects of the
general $tt^*$ geometry \doubref\CeVa\BCOV\ 
of the 2d SCFT on the string world-sheet. 

As for item iii), what is needed conceptually is 
to pass from the middle cohomology $H^3(X)$, which
enters the usual form of mirror symmetry, to a certain relative
cohomology group $H^3(X,\Z)$ defined by a submanifold $\Z\subset X$
associated to the background branes.
Accordingly, the r\^ole of the variation of 
Hodge structure in the construction of the usual
mirror map is taken over by the variation of the mixed 
Hodge structure on this relative cohomology group.
Together this leads to a nice correspondence between
the concepts of the fundamental 2d world-sheet theory and of the 
target space geometry, such as a geometric representation 
for the open-closed chiral ring. 

These concepts in TFT and geometry manifest themselves in the structure
of the $\cx N=1$ superpotential of the effective 
four-dimensional space-time theory. In particular, we will show how
the superpotential $W$ and its derivatives specify the
moduli dependent chiral ring $\Roc$, or, equivalently, the 
mixed Hodge structure on the relative cohomology group.

The basic topological data are a set of holomorphic 
potentials $\cx W_K(z_A)$ that are, in a quite precise sense, the
$\cx N=1$ counterparts of the holomorphic prepotential $\cx F(z_a)$ 
of $\cx N=2$ special geometry. They are among the building blocks
of the superpotential $W$ in the various flux and $D$-brane sectors
labeled by the index $K$. Above, the $z_A$ are the
scalar vev's of the chiral multiplets, both from the closed and 
open string sectors, which parametrize the 
$\cx N=1$ moduli space $\MM$.\foot{We will loosely refer to
the coupling space $\MM$ as a ``moduli space'', although in general there
will be a superpotential that lifts the flat directions. In particular
this makes sense for a scalar with a perturbatively flat
potential, as is the case for certain brane moduli.}
As will be discussed, on this space 
there exist flat, topological coordinates $t_A$ such that 
the derivatives 
$$
\fc{\p}{\p t_A}\fc{\p}{\p t_B}\, \cx W_K(t_\dd) = C_{AB}^{\ \ K}(t_\dd)$$
represent the structure constants $C_{AB}^{\ \ K}$
of the open-closed chiral ring $\Roc$. The special coordinates $t_A$
are defined by an integrable connection $\nabla$, that defines a
system of differential equations of Picard-Fuchs type for the
superpotential. The instanton corrected superpotential is a solution
of this system and can be determined by standard methods.

Since the mathematical details are a little involved, we will
present in this note some of the basic ideas and results,
and defer more thorough arguments  and computations
to a companion paper \copa. In Appendix A we include
a brief summary of the basic concepts in 
2d TFT, such as the chiral ring, 
and their connection to ordinary special geometry. 
This section will be useful for reference and
provides some background material for our
discussion of $\cx N=1$ special geometry.

\newsec{$\cx N=1$ Superpotentials in type II \CY compactifications}

We begin the discussion of the holomorphic $\cx N=1$ special geometry 
with a physical characterization of an object that will be 
central to much of the following: this is the relative period 
vector $\Pi^\Si$ which 
encodes the instanton corrected $\cx N=1$
superpotential of the open-closed string theory.

We consider
a  compactification of the closed 
type IIB string on a \CY 3-fold $X$ with Hodge numbers 
$h^{p,q}=$ dim $H^{p,q}$. The effective theory at low energies 
is an $\cx N=2$ supergravity with a generic Abelian gauge group 
$U(1)^{h^{1,2}+1}$.
There are two closely related modifications of this $\cx N=2$
supersymmetric  type IIB
background that break supersymmetry in a way that may be described
by an effective $\cx N=1$ supergravity action. 

The first modification is a deformation in 
the closed string theory, obtained by adding background
fluxes $H=H^{RR}+\tau H^{NS}$ of the 2-form gauge fields on $X$.
This leads to an $\cx N=1$ superpotential of the form 
\refs{\Guk,\TV,\PMsusy}:
\eqn\cssp{
W_{cl}(z_a)= \int \Om \wedge H \,=\, \sum_\al N_\al\cdot\Pi^\al(z_a) .}
This superpotential depends on the vev's of certain 
scalars $z_a$ in 
$\cx N=1$ chiral multiplets, 
and these vev's represent the complex structure 
deformations from the closed string sector.
The moduli dependence is encoded in 
the period vector of the holomorphic $(3,0)$-form $\Om$ on $X$:
$$
\Pi^\al= \int_{\Ga^\al} \Om(z_a),\qquad \Ga^\al\in H_3(X,\ZZ).
$$
The parameters $N_\al$ in \cssp\ specify the integer 
3-form fluxes on $X$.  That is, 
$N_\al = n_\al +\tau m_\al$, where the $n_\al, m_\al$ are integers
and $\tau$ is the type IIB string  dilaton \TV.

The second supersymmetry-breaking modification is to 
introduce an open string sector by adding
background (D-)branes that wrap 
supersymmetric cycles $\C_\nu\in H_{2n}(X)$ and simultaneously 
fill space-time.
The $\cx N=1$ superpotential for these branes is computed by the
holomorphic Chern-Simons functional \WitCS.
Specifically, we will consider 5-branes  wrapped on a set of 2-cycles 
$\{\C_\nu\}$, for 
which the superpotential is given by \refs{\AVi,\WQCD,\KKLM}:
\eqn\ossp{
W_{op}(z_a,\zh_k)\ =\ 
N_\nu \cdot \int_{\Gah^\nu} \Om(z_a)\ 
=\ \sum_\nu N_\nu \cdot \Pi^\nu(z_a,\zh_\al) \,.
}
Here $\Gah^\nu$ denotes a special Lagrangian  3-chain with boundary 
$\p\Gah^\nu \supset \C_\nu$  and the $\zh_\al$ are the brane moduli
from the open string sector. Moreover $N_\nu= n_\nu+\tau m_\nu$,
where $n_\nu$ $(m_\nu)$ are the numbers of D5 
(NS5) branes. The moduli $\zh_\al$
parametrize the position of the D-branes in $X$ in a special
way, namely by measuring the volumes of the 3-chains $\Gah^\nu$
whose boundaries are wrapped by the D-branes. The precise definition
of the good open string moduli will be one of the
outcomes of the following discussion. 

It is natural to consider a combination of the two types of 
supersymmetry breaking backgrounds, and to study the general 
superpotential on the full deformation space $\MM$ parametrized
by  the closed and open string moduli, $z_a$ and $\zh_\al$, respectively. 
In the low energy effective action, the two 
contributions  combine into the section of a single line bundle $\cx L$
over $\MM$, 
and are really on the same level. Note that the line bundle $\cx L$ is
identified with the bundle of holomorphic $(3,0)$ forms on $X$. 
Also, in string theory, the 
distinction between  RR fluxes and background D-branes is 
somewhat ambiguous: In the large $N$ transition of \CVlargeN\ a 
type IIB background with $N$ D-branes is replaced by
a type IIB background on a different manifold $X'$ without 
branes but $N$ units of flux. 

We thus consider the most general superpotential from background 
fluxes and branes of the form
\eqn\gensp{
W_{\cx N=1}\ =\ W_{cl}(z_a)+W_{op}(z_a,\zh_\al)\ =\ 
\sum_\Si N_\Si \,\cdot \, \Pi^\Si(z_A),
}
where $\Pi^\Si$  is the {\it relative period vector}\foot{More precisely,
the pairing is defined in relative (co-)homology, as discussed in 
sect. 5.}
\eqn\lpv{ \Pi^\Si(z_A) = \int_{\Ga^\Si} 
\Om\,, \qquad \Ga^\Si \in H_3(X,\Y,\ZZ),\qquad z_A\equiv\{z_a,\zh_\al\}. }
Here $ H_3(X,\Y)$ is the group of relative homology cycles 
on $X$ with boundaries on $\Y=\cup_\nu \C_\nu$, where 
$\C_\nu \subset \p\Gah^\nu$. Its
elements are $i)$ the 
familiar 3-cycles without boundaries, whose volumes
specify the flux superpotential and $ii)$ the 3-chains
with boundaries on the 2-cycles $B_\nu$. 
The vector $\Pi^\Si$ thus uniformly
combines the period 
integrals of $\Om$ over 3-cycles $\Gamma^\al$, with 
integrals over 3-chains $\Gah^\nu$ whose boundaries are wrapped by D-branes. 

In the following we will show that, for appropriate normalization of $\Om$,
the relative period vector is of the form
\eqn\permi{
\Pi^\Si=(1,t_A,\cx W_K,...)\ ,
}
where $t_A$ are flat topological coordinates on $\MM$ and $\cx W_K$ the
{\it holomorphic potentials of $\cx N=1$ special geometry}. The latter 
determine the ring structure constants $C_{AB}^{\ \ K}$
of the extended open-closed chiral ring $\Roc$
$$
\phix 1_A \cdot \phix 1 _B = C_{AB}^{\ \ K}\ \phix 2 _K.
$$
Here the $\phi_I^{(q)}$ denote a basis of 2d superfields that span the 
local BRST cohomology of the topological sector of the 2d 
world-sheet theory for the compactification on $X$.\foot{See sect. 4
and App. A for a further discussion and references.}
The ring structure constants are given in terms of the derivatives 
\eqn\oscrscs{
C_{AB}^{\ \ K}(t_\dd)=\fc{\p}{\p t_A}\fc{\p}{\p t_B}\cx W_K(t_\dd).
}
This formula is the $\cx N=1$ counterpart of the 
well-known equation (A.3) that describes the 
chiral ring constants of $\cx N=2$ special geometry in terms of 
the prepotential $\cx F$. 

In the next two sections
we will outline the relation between the 
relative period vector $\Pi^\Si$ that defines
the holomorphic $\cx N=1$ superpotential \gensp,
and the chiral ring of the underlying TFT on the string
world-sheet.  Concretely,  the elements $\phi_A$ 
of the chiral ring $\Roc$
have a geometric representation in the B-model as elements of a certain
relative cohomology group $H^3(X,\Z)$.\foot{Here $\Z$ denotes a 
union of hypersurfaces in $X$ passing through the cycles $B_\nu$,
as discussed below.} On the bundle of open-closed
B-models over $\MM$, there exists a topological {\it flat} connection $\nabla$,
which is the Gauss-Manin connection on the relative cohomology bundle.
The flatness of $\nabla$ predicts the existence of 
special topological coordinates $t_A$, for which the covariant 
derivatives $\nabla_A$ reduce to 
the ordinary derivatives $\p_A=\fc{\p}{\p t_A}$; these are precisely
the special coordinates on the $\cx N=1$ moduli space $\MM$
that appear in \permi.
Moreover, the geometric representation of the chiral ring
$\Roc$ leads to an exact
expression for the holomorphic potentials $\cx W_K$ in terms of the 
``relative period matrix'' for $H^3(X,\Z)$. This matrix describes 
the projection of the moduli dependent chiral ring onto a fixed, 
topological field basis and represents the counterpart of 
the familiar period matrix (A.4) in $\cx N=1$ special geometry.
\vskip 3pt

Before we turn to this discussion, 
it is worth to comment on 
the physical meaning of this
extra structure in the effective four-dimensional supergravity 
theory. This is important,  as it distinguishes the 
string effective theory from a generic $\cx N=1$ supergravity. 

Recall that the type IIB string
has 5-branes in the NS and the RR sectors, which may be 
wrapped on supersymmetric 3-cycles in $H_3(X,B)$. These states
are interpreted as domain walls in four dimensions with non-zero BPS 
charge \doubref\WQCD\GVW. We 
are thus led to identify the integral 
cohomology lattice dual to that in \lpv, 
with the lattice of four-dimensional BPS charges\foot{with two 
copies for the RR and NS sector, respectively}

\eqn\bpslat{
\Gamma^{4d}_{BPS}=H^3(X,B;\ZZ)\otimes(1\oplus \tau).}

\noi
The BPS tension of a domain wall with charge $Q\in \Gamma_{BPS}$
is then determined by the volume of the wrapped 3-manifold, as measured by 
the relative period vector $\Pi^\Si$. In fact 
the tension of the domain wall is proportional to a change in the $\cx N=1$ 
superpotential on the two sides of the domain wall \doubref\GVW\TV. Thus
the space-time interpretation of the holomorphic
$\cx N=1$ special geometry of 
the string effective supergravity is in terms of the BPS geometry 
of the 4d domain walls with charges $Q\in\Ga_{BPS}$, very much as 
the $\cx N=2$ special geometry describes the more familiar 
BPS geometry of four-dimensional particles. 

One may also interpret the $\cx N=1$ special geometry 
in terms of masses of BPS particles, by compactifying on a further $T^2$ 
to a two-dimensional theory with $\cx N=2$ supersymmetry, with the
5-branes wrapped on the extra $T^2$. The same
type of theories also arises in a \CY 4-fold compactification of the
type IIB string, where the BPS particles are represented by 5-branes
wrapped on special Lagrangian 4-cycles in the 4-fold. In some
cases such a closed string 4-fold compactification is dual to the
open-closed type II 
string compactification on the 3-fold $X$ times the extra $T^2$ \PM.
This then provides an identification of the
presently discussed $\cx N=1$ BPS geometry of 4d domain walls 
on the 3-fold with $D$-branes, with the
2d $\cx N=2$ geometry of 
BPS {\it states} in the 4-fold compactification.

Another interesting aspect is that 
the ring structure constants enjoy an integral instanton
expansion \OV:
\eqn\instexpan{
C_{AB}^{(inst)  \ K}\ =\
\sum_{\{n_C\}}\sum_k n_A n_B N^{(K)}_{n_1\dots n_M}
\prod_C e^{2\pi i k n_Ct_C},} 
where $M={\rm dim}\MM$ and the coefficients $N_{\dd}^{(K)}$ are integers that 
count the number of certain BPS multiplets of the theory compactified
to two dimensions, in the topological sector
labeled by $K$. Alternatively, these numbers may be thought of 
as counting the
(appropriately defined) number of world-sheet instantons of 
sphere or disc topologies.\foot{In the case
of sphere topologies, the above multi-covering formula is again
defined in a two-dimensional compactification; see \PM\ for
a discussion.} 

Most importantly, the instanton expansion 
\instexpan\ can in many
cases, if not in all, be directly identified with  
genuine {\it space-time} instanton corrections to the superpotential
in the RR sector. The reason is the familiar fact that the four-dimensional
coupling constants in the RR sector are not determined by the 
four-dimensional string coupling, but rather by the geometric moduli $t_A$.
The
statement that the space-time instanton expansion
has integrality
properties, attributable to the counting of
BPS states, is quite remarkable and
distinguishes this class of effective supergravities from generic ones. 

\newsec{Open-closed chiral ring and relative cohomology}
As reviewed in App. A, the elements of the 
chiral ring of the closed string B-model on $X$ have
a geometric representation as the cohomology elements
in $H^3(X)$. Moreover the gradation by $U(1)$ charge 
of the chiral ring corresponds to the Hodge decomposition
$H^3(X)=\oplus_p\,  H^{p,\, 3-p}(X)$.
We will now describe a similar
geometric representation of the chiral ring of the open-closed B-model,
in terms of a relative cohomology group $H^3(X,\Z)$.\foot{For an introduction
to relative cohomology, see e.g. \Kar.}

The open-closed chiral ring is an extension of the 
closed string chiral ring, and combines operators
from both the bulk and boundary sectors.
Geometrically, the new structure from the open string
sector is the submanifold $B\subset X$ wrapped 
by the D-branes.\foot{We will restrict our discussion to trivial
line bundles on the two-cycle $B_\nu$.} 
Since the bulk sector of
the closed string is represented by $H^3(X)$, the
open-closed chiral ring should correspond to an
extension of this group by new elements originating
in the open string sector on $B$.\foot{More precisely,
the ring multiplication for the bulk operators could
be a priori different on closed and open string world-sheets.}
It is natural to expect that this extension is 
simply the dual of the space $H_3(X,B)$ which 
is underlying the superpotential \gensp\ from the fluxes
and branes.
The dual space is the relative cohomology group $H^3(X,B)$,
and we will see that it gives indeed the right answer.

Let us discuss at this point 
more generally a relative cohomology group $H^3(X,\Z)$,
where $\Z=B$ for the moment;
the motivation for this is that 
we will eventually give an argument
that allows to replace the boundary $B$ by a simpler object.

The relative cohomology group $H^3(X,\Z)$ fits into a long
exact sequence 
\eqn\lseq{
...\to H^2(\Z) \to H^3(X,\Z) \to H^3(X) \to ...
}
To simplify the discussion, let us assume 
that the maps on the left and right hand side 
are trivial. This assumption can be easily checked in practice 
and is justified for a large class of D-brane geometries \copa.
 In this case the group $H^3(X,\Z)$ is essentially
an extension of a space which combines the 3-forms on $X$ and
the 2-forms on $\Z\subset X$. The elements of  $H^3(X)$ have
an obvious interpretation as describing the (topological) 
closed string states in the open-closed string compactification\foot{In
2d CFT language, these are bulk operators inserted in the interior of
a world-sheet with boundary.}.
On the other hand the group $H^2(\Z)$ will describe the new degrees
of freedom originating in the open string sector.

An element $\vx \Theta \in H^3(X,\Z)$ may be 
specified by a pair of differential forms 
$$
\vx \Theta = (\Theta_X,\theta_\Z),
$$
where $\Theta_X$ is a 3-form on $X$ and $\theta_\Z$ a 2-form
on $\Z$. The differential is $$d\vx \Theta = (d \Theta_X,
i^*\Theta_X-d \theta_\Z),$$ where $i^*$ is 
the map on forms deriving from the embedding $i:\, \Z\to X$.
Thus $d\vx \Theta =0$ implies that $\Theta_X$ is closed on $X$  
and restricts to an exact form on $\Z$. Moreover
the equivalence relation is 
\eqn\relc{
\vx \Theta \sim \vx \Theta + (d \om,i^*\om-d\phi),}
where $\om$ ($\phi$) is a 2-form on $X$ (1-form on $\Z$). 
Note that this equation says that the exact form $-d\om$ on $X$
is not necessarily trivial in $H^3(X,\Z)$, but equivalent to the 2-form
$i^*\om$ on $\Z$. Thus a form that represents a trivial
operator in the closed string theory, such as an exact 
piece of the holomorphic $(3,0)$ form, may
lead to non-trivial elements in the open string extension
of the chiral ring -- specifically from ``boundary terms'' on the 
submanifold $\Z\subset X$. 

Our aim is to connect the moduli dependence
of the chiral ring $\Roc$ for a family of geometries parametrized
by the couplings $z_A$, to the superpotential $W$ and its derivatives.
Geometrically, the coordinates $z_A=(z_a,\zh_\al)$ on the moduli space $\MM$ 
describe the complex structure of the
manifold $X$ (closed string sector) and the ``location'' of 
the D-branes specified by the map $\Z \hookrightarrow X$ (open string
sector), respectively. The relative cohomology groups $H^3(X,\Z)$ fit together
to a bundle $\cx V$ over $\MM$ that may be identified with the bundle
of RR ground states $\bra I(z_A,\bb z_{\bb A})$ for the B-model on the
family of D-brane geometries parametrized by the couplings $z_A$.
Here we are using the correspondence between the 
chiral ring elements $\phi_I$
and ground states $\bra I$ in the RR sector; specifically the
ground state $\bra I$ may be obtained from the canonical vacuum $\bra 0$
by inserting the operator $\phi_I$ in the twisted 
path integral on a world-sheet with boundary \BCOV.

In the following we discuss the case of a single brane wrapped
on a 2-cycle $B$; the superpotential for a compactification with 
several non-intersecting 
branes wrapped on a collection of cycles $\{B_\nu\}$ 
is the sum of the individual superpotentials and can be treated similarly.
As is well-known, a supersymmetric configuration of a D-brane wrapped
on a 2-cycle $B\in X$ requires $B$ to be holomorphic \SSTwo.
On general grounds, the condition for supersymmetry in field theory
is of the form $W'(z_A)=0$, where the prime denotes
an arbitrary derivative in the moduli. A non-trivial superpotential 
in the moduli $z_A$ is thus defined on a family $\cx B$ of 2-cycles,
whose members are in general non-holomorphic, except at those values of $z_A$
where $W$ has critical points.

A study of the relative cohomology group $H^3(X,B)$ for $B\in \cx B$ 
would thus involve in general non-holomorphic maps $B\hookrightarrow X$.
To avoid such a complication, we will now give an argument that
allows to replace the group $H^3(X,B)$ by another relative cohomology
group $H^3(X,\Z)$, where $\Z$ is a member of a 
family of holomorphic hypersurfaces in 
$X$ that pass through the boundary cycles in $\cx B$.%
\foot{Simply speaking, the possibility to 
replace the group $H^3(X,B)$ by $H^3(X,\Z)$ is due the fact that
the topological B-model depends only on the complex structure
type of moduli, not on the K\"ahler type. Thus, in general there may be
a whole family of deformations of the 2-cycle $\Gamma_2$,
whose members all have the same superpotential. In the situation
discussed below, the group $H^3(X,\Z)$, as
defined by an appropriate holomorphic hypersurface $\Z$ 
passing through this family, is a good object for capturing  
the superpotential. See also 
\Grifdeq\ for a related discussion.}

The superpotential for the brane wrapped on $B$ is proportional
to the volume \ossp\ of a minimal volume 3-chain $\Ga$ with boundary
$\p\Ga\supset B$. One way to solve the minimal volume condition 
is to slice $\Ga$ into 2-cycles along a path $I$. This can be achieved 
by intersecting $\Gamma$ with a family\foot{For simplicity, 
we will consider here only one-parameter families, 
corresponding to brane wrappings with a single modulus.}
of holomorphic hypersurfaces $\Z(\z)$, 
where $\z$ is a complex parameter. The intersection
of the hypersurface $\Z(\z)$ with $\Gamma$ is a family of 2-cycles 
$\Gamma_2(\z)$ of minimal volume $V(\z)$, and 
the integral \lpv\ can be written as
\eqn\wgamma{
W_\Gamma= \int_\Gamma \Om = \int_{\z_0}^{\z_1}V(\z)d\z.
}
Here the path $I$ in the $\z$-plane is determined by the
minimal volume condition for $\Gamma$. The existence of the
appropriate family of hypersurfaces is the assumption 
used in the following discussion.\foot{This is a relatively mild
assumption that is in particular satisfied for a large
class of D-brane geometries \copa. The above idea 
was used in \KLMVW\ in the context of (mirrors of) ALE fibrations, 
which give a natural slicing of a chain $\Ga$ into a 1-parameter family 
of 2-cycles in the ALE fiber parametrized by a path on the base.}

The interval in the $\z$-plane ends at a
specific value, say $\z=\hx z$, for which
the hypersurface $\Z\equiv \Z(\hx z)$ passes through the 
boundary 2-cycle $B$ which is wrapped by the D5-brane.
Varying the position of the D-brane leads to the variation 
of the brane superpotential:
\eqn\delw{
\delta W_{\hx\Gamma}\ \sim\  V(\hx z)\ =\ 
\int_{\Gamma_2(\hx z)} \om,
}
where $\om$ is a 2-form on $\Z$. 
As discussed in more detail in \copa, the form $\om$ is a
holomorphic $(2,0)$ form on $\Z$ obtained from a Poincar\'e residue of $\Om$. 
Note that $\delta W_\Gamma$ vanishes if $\Gamma_2$ is holomorphic 
\refs{\WQCD,\AVi,\KKLM}, as expected.

By the above construction, the open string extension of the chiral
ring is represented by the holomorphic 2-form $\om$ on the 
hypersurface $\Z$.
As shown in \copa, products of chiral fields in the bulk
with $\om$ generate additional elements in $\Roc$ 
which can be identified with 2-forms of lower 
holomorphic degree in $H^2(\Z)$. These elements 
in $H^2(\Z)$, together with the elements in $H^3(X)$
from the closed string sector, combine into the
relative cohomology group $H^3(X,\Z)$, as described by the
sequence \lseq.

\newsec{Exact 4d 
$\cx N=1$ superpotential from the integrable connection $\nabla$}
The relative cohomology bundle $\cx V$ over $\MM$ with fibers 
$H^3(X,\Z)$ comes with a structure
that comprises all the ingredients needed
to define the generalized special geometry of the open-closed B-model.
This is the family of mixed Hodge structures \Del\ on the bundle $\cx V$
over $\MM$, and a flat Gauss-Manin
connection $\nabla$ defined on it.
We will be brief in the following, leaving more detailed explanations
and computations to \copa. Here we sketch how the
integrable connection $\nabla$ on $\cx V$ determines a
set of flat topological coordinates on the $\cx N=1$ moduli space $\MM$
and the exact instanton corrected $\cx N=1$ superpotential of the effective
4d theory. 
The following discussion will be in many respects be similar to the one in the
Appendix,  where we review how the flat connection
for the moduli dependent closed string chiral ring leads to the 
exact formula for the $\cx N=2$ prepotential~$\cx F$. 

Specifically we want to relate the holomorphic potentials
$\cx W_K$ to the moduli dependence of the structure constants of
the open-closed chiral ring, $\Roc$.
For this purpose, we would like to know at each point in the coupling space 
$z_A\in \MM$, the basis $\{\vx \Phi_I\}$ for $H^3(X,\Z)$ that represents
the elements $\phi_I$ of the chiral ring, or equivalently,
the RR ground states $\bra I$ created by the 
operators $\phi_I$ from a canonical vacuum $\bra 0$.

An important structure in the TFT is the existence
of a preserved $U(1)$ charge. This gives an integral grading to the space
$V$ of RR ground states 
$$
V=\oplus_q V^{(q)}.
$$
Special geometry essentially describes the position of the 
subspaces $V^{(q)}$ in $V$, defined relative to a constant basis for $V$.

At grade zero, there is a unique element of the chiral ring corresponding
to the unit operator $\phix 0$ that flows to the canonical vacuum in
the RR sector,~$\bra 0$.\foot{The $U(1)$ charge of $\bra 0$ is shifted
by $-\hx c/2$ under the spectral flow from the NSNS to the RR sector;
we will refer to integral grades without this shift also in 
the RR sector.}

The most interesting sector is that of the fields $\phix 1 _A$
of grade one, which represent the 
deformations of the 2d TFT parametrized
by the couplings $z_A\in \MM$. In the present context,
these fields generate the full chiral ring. 
For example, grade two fields can be generated by the OPE:
\eqn\gradetwo{
\phix 1 _A (z_\dd)\cdot \phix 1 _B(z_\dd)= C_{AB}^{\ \ K}(z_\dd)\, 
\phix 2 _K(z_\dd).
}
The moduli dependence of a chiral field
$\phix q _I(z_\dd)$ can be described by projecting to a fixed, constant
basis for $V$. Such a basis $\{\vx \Ga_\Si\}$ may be defined
as the dual of a basis $\Ga^\Si$ of topological cycles for
the relative homology group $H_3(X,\Z)$. The transition
to the constant basis is
$$
\Phiv q _I(z_\dd) = \Pi_I^\Si(z_\dd)\, \cdot \,  \vx \Ga_\Si\, .
$$
Here $\Pi_I^\Si(z_\dd)$ is the {\it relative period matrix} for the relative
cohomology group $H^3(X,\Z)$:
\eqn\relpair{
\Pi_I^\Si= \langle \vx \Phi_I,\Ga^\Si\rangle=
\int_{\Ga^\Si} \Phi_i - \int_{\p \Gamma^\Si}\phi_I,
}
for $\vx \Phi_I=(\Phi_I,\phi_I)\in H^3(X,\Z)$. The
linear combination of integrals on the
r.h.s. of the above equation defines the dual pairing 
in relative (co-)homology, invariant under the equivalence relation \relc.

The grade one elements $\Phiv 1 _A$ can be identified by the 
fundamental property
$$
(\nabla_A-C_A)\,\bra I = 0,
$$
which expresses the fact that
an insertion of the grade one operator $\phix 1 _A$ 
(represented by the matrix $C_A$) in the path integral
is equivalent to taking a derivative in the 
$A$ direction. The connection terms in $\nabla_A$ depend on the choice
of coordinates $z_A$ on $\MM$ and are not determined by this argument.
However, $\nabla_A$ is a {\it flat} connection on the
bundle $\cx V$ over $\MM$ with fibers $V$, namely the
Gauss-Manin connection on the relative cohomology bundle $\cx V \sim 
H^3(X,\Z)\otimes \cx O_{\MM}$.

In particular, the flatness of the connection $\nabla$ implies
the existence of flat coordinates $t_A(z_\dd)$ for which the 
covariant derivatives reduce to ordinary ones, 
$\nabla_A \, \to\,  \p_A=\p/\p t_A$. 
In these coordinates the relative period matrix satisfies  the
following system of linear differential equations:
\eqn\lde{
\Big(\fc{\p}{\p t_A}-C_A(t_\dd)\Big)\, \Pi_I^\Si(t_\dd)=0.
}
If we order the basis $\{\Phiv q _I\}$ by increasing grade $q$,
these equations imply that $\Pi_I^\Si$ can be put into 
upper block triangular form 
with constant entries on the block diagonals. In fact
one may chose $\Phiv 0 = (\Om,0)$ as a basis element for this
space \copa, such that the first row  
$$
\Pi^\Si\equiv \Pi_0^\Si = \int_{\Ga^\Si} \Om
$$
describes the periods of the holomorphic $(3,0)$ form $\Om$ on
the basis $\{\Ga_\Si\}$ of topological 3-cycles and 3-chains for 
$H_3(X,\Z)$. 

At grade one, the content of the equations \lde\ is the definition
of the flat coordinates $t_A(z_\dd)$:
\eqn\mmp{
t_A(z) = \fc{\Pi_0^A}{\Pi_0^0}\ =\ \fc{\int_{\Ga^A}\Om}{\int_{\Ga^0} \Om},}
for $A=1,...,\rmx{dim}(\MM)$.
This is the {\it $\cx N=1$  mirror map}
for the $\cx N=1$ chiral multiplets, which
defines the flat coordinates on the deformation space $\MM$ as the ratio
of certain period and chain integrals on the manifold $X$ \PM.  
A different definition
of the open string moduli, namely in terms of the BPS tension of 
4d domain walls, had been first given in \doubref\OV\AVi; it
agrees with the above definition of flat topological coordinates, 
at least for the cases studied so far.

In the flat coordinates $t_A$, the relative period matrix has the general form 
\eqn\rpm{
\pmatrix{
\vbox{\tabskip=10pt\halign{\strut
#&\hfil~$#$~\hfil\vrule&\hfil~$#$~\hfil\vrule&\hfil~$#$~\hfil\vrule
&\hfil~$#$~\hfil\vrule~\vrule&\hskip 12pt $#$~\hfil\cr
&1&t_A&\cal{W}_K&...&(q=0)\cr
\noalign{\hrule}
&0&\delta_B^A&\partial_B\cal{W}_K&...&(q=1)\cr
\noalign{\hrule}
&0&0&\eta_L^K&...&(q=2)\cr
}}}}
where $\eta_L^K$ is some constant matrix. 
At grade two we then get from \lde\ and \rpm:
$$
\fc{\p}{\p t_A}\fc{\p}{\p t_B} \cx W_K(z_\dd) = C_{AB}^{\ \ K}(z_\dd)
$$
This is the promised relation that 
expresses the ring structure constants in terms of
the holomorphic potentials
\eqn\holp{
\cx W_K(z_\dd) = \int_{\Ga^K}\Om.
}
Inverting the mirror map \mmp, and inserting the result into
the potentials \holp, one then finally obtains the instanton
expansion \instexpan\ of the superpotentials.

The linear system \lde\ terminates at grade $q\leq 3$. 
Eliminating the lower rows in this system in favor of
the first row, namely the relative period vector 
$\Pi^\Si$, one obtains a system of coupled, higher order
differential equations
\eqn\pifu{
\cx L_A \ \Pi^\Si(z_\dd)=0.}
These equations comprise a Picard-Fuchs system for  
the of the relative cohomology group $H^3(X,\Z)$.
Solutions to the above 
differential operators, most notably
expansions around the classical point $z_A=0$, 
can then easily be obtained 
by standard methods, as has been exemplified in \doubref\PM\LM.

This then gives a very effective way to determine
the holomorphic potentials $\cx W_K(t_\dd)$,
very similar to how $\cx N=2$ prepotentials were computed in the past.
Via \gensp, these expansions represent non-perturbatively exact
space-time instanton contributions to the $\cx N=1$
superpotential $W(t_\dd)$ of the four-dimensional string effective
supergravity theory.

Although many of the above equations will have looked familiar
to the reader who is acquainted
with the connection between 2d TFT
and $\cx N=2$ special geometry, the validity of these
equations in the present, $\cx N=1$ supersymmetric 
context is very non-trivial. As alluded to before,
the underlying mathematical structure is the mixed Hodge
structure on the relative cohomology bundle $\cx V$,
and the Gauss-Manin connection defined on it.
The precise arguments why
these mathematical concepts can be identified with the 
TFT concepts discussed here, leading to the equations
presented in this section, will be given in ref.~\copa.

\newsec{Conclusion and outlook}

Summarizing, what we have outlined in the present letter is a general
geometric structure that underlies the $\cx N=1$ superpotentials of certain
flux and brane configurations in type II string compactifications
on Calabi-Yau threefolds. This holomorphic $\cx N=1$ special geometry is in
many respects similar to (and in a sense an extension of)
the special geometry of $\cx N=2$ supersymmetric type II string
compactifications. It puts the computation of the instanton-corrected 
superpotential $W$ of the four-dimensional $\cx N=1$ string effective
action on a very similar footing as the familiar
computation of the $\cx N=2$ prepotential $\cx F$ via mirror symmetry.

As already mentioned, the purpose of this letter is to give an
overview of these of our main ideas and results, 
which will be further developed and explained in more detail in the
companion paper \copa. 
We will in particular show there how these concepts 
can be made very explicit in
the framework of linear sigma models, or toric geometry. 
For this broad class of toric geometries, 
$\cx N=1$ special geometry 
provides an efficient and systematic method to compute the 
instanton corrected superpotentials $W$. 

Specifically, we will derive a simple toric representation
of the open-closed chiral ring in terms of the relative cohomology,
and of a complete system of GKZ type differential equations.
In passing we will 
also show that for non-compact threefolds, the degree of the slicing 
hypersurface \Z\ is a free parameter and gives a natural geometric 
interpretation to the so-called ``framing ambiguity'', a
discrete quantum number in the open string sector discovered in  
ref.\AVii.

Finally we like to recall that, as discussed in 
sect.~2, the space-time instanton expansion
\instexpan\ of $W$ in the string effective theory 
has certain integrality properties, leading to a 
highly distinguished class of $\cx N=1$ supergravities.
It is interesting to ask whether the integrality of the 
expansion can be directly linked to the holomorphic special 
geometry discussed in this paper.
Indeed both structures point to some 
non-perturbative ``duality group'' acting on the coupling space $\MM$ of the 
string compactification,
which is likely to originate from the monodromy group of the solutions of 
the differential equations. 
The superpotential $W$
must then transform properly under $S$ transformations.
For appropriate groups $S$, these ``automorphic'' 
functions often have miraculous integrality properties,
like for example the famous $j$-function for the group $SL(2,\ZZ)$. 
The idea that the 
effective theory might be largely constrained by a non-perturbative
duality group has been discussed already in the string phenomenology of
the twentieth century; 
there the strategy was to assume a -- simple enough -- 
duality group, and then to attempt to determine the superpotential 
by identifying the appropriate automorphic functions (see e.g. \auto).

We are here in a somewhat
opposite situation, where we can compute the automorphic function,
and in principle derive also the duality group,
from a system of differential equations. If an open-closed string
compactification has an interesting enough duality group,
this may lead to far reaching constraints on other
quantities in the effective action, such as the K\"ahler potential
for the moduli of the B-model. Clearly it would be interesting
to study this further, perhaps in the context of the $tt^*$ equations
of refs.\doubref\CeVa\BCOV.
\vskip 6pt

\goodbreak
\noi {\bf Acknowledgments:} 
We thank Calin Lazaroiu and Chris Peters for discussions.
This work was supported in part by the DFG
and by funds provided by the DOE under grant number DE-FG03-84ER-40168. 
WL would like to thank the USC-CIT center in Los Angeles
for hospitality; moreover
WL and NW would like to thank 
the Isaac Newton Institute in Cambridge for hospitality.

\appendix{A}{$\cx N=2$ special geometry and 2d TFT}
Here we give a concise summary of the relation between
the $\cx N=2$ special geometry and the 2d TFT on the
string world-sheet; for a thorough discussion of this
connection, we refer to ref.\BCOV. 

The basic topological datum of the $\cx N=2$ supergravity theory
is the holomorphic prepotential $\cx F(z_a)$, where $z_a$ are
some coordinates on the vector moduli space $\Mm$. The 2d interpretation
of the prepotential $\cx F(z_a)$ in the TFT is as a generating function
for the structure constants of the chiral ring $\cx R$~\LVW:
\eqn\cscr{
\phi^{(q)}_i \cdot \phi^{(q')}_j = C_{ij}^{\ k} \phi^{(q+q')}_k.}
Here the $\phi_i^{(q)}$ denote a basis of 2d superfields that span the 
local BRST cohomology of the TFT. The superscript denotes 
the charge $q=(q_L,q_R)$ of a field under the left- and right-moving 
$U(1)$'s of the super-conformal algebra on the world-sheet.
The fields $\phi_a$ of $U(1)$ charges $(1,-1)$,
 labeled by subscripts
from the beginning of the alphabet, represent the marginal deformations of 
the 2d field theory and are in one-to-one correspondence with the moduli 
$z_a$. 

By spectral flow, the 
operators $\phi_i$ are in one-to-one correspondence with the 2d ground states $\bra i$
in the RR sector.  The topological  RR ground states
can be generated from a canonical vacuum $\bra 0$ by 
inserting fields $\phi_i$ in the twisted path integral, 
which leads to a 
representation of the form: $\bra i = \phi_i \bra 0$.
An important property of the TFT is the existence of a
{\it flat} topological connection $\nabla$ on the bundle $\cx V$ of 
RR ground states $\bra {i (z_a,\bb z _{\bb a})}$ 
over $\Mm$.
By a familiar argument, which identifies a derivative with
respect to the deformation parameter with an insertion of the operator
in the path integral, one obtains the relation:
\eqn\flatco{
D_a \bra i = (\nabla_a-C_a)\, \bra i = 0.}
Here $C_a$ is interpreted as a matrix representing the multiplication
with the field $\phi_a$. The flatness of the connection $\nabla$ and the
precise form of the  connection terms in \flatco\ have been studied 
in \BCOV; in 
particular the object $D_a$ is the $tt^*$ connection
of ref.\CeVa.

The basic relation connecting the chiral ring coefficients $C_a$ with the 
$\cx N=2$ prepotential $\cx F$ in the supergravity is 
\eqn\cscrii{
C_{ab}^{\ \ c}= \p_a\p_b\p_c \cx F(t_\dd),\qquad \p_a= \fc{\p}{\p t_a}.
}
Here we have introduced the topological flat coordinates $t_a$, which 
are the local deformation parameters in the 2d world-sheet action 
such that the connection terms vanish, i.e., $\nabla_a \to \p_a$.

The TFT concepts we were just discussing
have an explicit geometric realization
in the topological B-model, in terms of the family of Hodge structures 
on the middle cohomology $H^3(X,\CC)$. The elements $\phi^{(q)}_i$ 
of the chiral ring are
identified with elements of the Hodge spaces $H^{3-q,q}(X)$. 
The unique element $\phi^{(0)}$ of zero $U(1)$ charge
is identified with the unique holomorphic $(3,0)$ form $\Om$ on $X$.
Moreover, deformations arising from the charge one fields are identified
with the deformations of the complex structures on $X$. 
The spaces $H^3(X,\CC)$ fit together to a locally trivial
bundle $\cx V$ over $\Mm$ which admits a flat connection $\nabla$, the
so-called Gauss-Manin connection. As the complex structure
varies with the moduli $z_a$, the definition of a $(p,q)$ form changes 
and therefore the bundles with fiber $H^{3-q,q}(X)$ are non-trivial.
The moduli dependence of a representative $\Phi^{(q)}_i(z_\dd)
\in H^{3-q,q}(X)$ 
for the field $\phi_i^{(q)}$ of definite $U(1)$ charge, 
can be specified by its projection onto 
a fixed, moduli independent basis $\{\Ga_\al\} \in H^3(X)$.
The transition functions are summarized in the period matrix:
\eqn\gsii{
\Pi_i^\al(z_\dd)=\
\langle i \bra \al = \int_{\Ga^{\al}} \Phi^{(q)}_i(z_\dd),
\qquad \Ga^{\al}\in H_3(X,\ZZ) \,. }
Here $\Ga^\al\in H_3(X,\ZZ)$ is a basis dual to the 
constant basis $\Ga_\al$. 
{}From \flatco\ we see that the period matrix satisfies the differential equation 
\eqn\flatii{
D_a \Pi_i^\al(z_\dd) = (\nabla_a-C_a(z_\dd))\, \Pi_i^\al(z_{\dd}) = 0.}

Let us order the basis $\{\Phi_i^{(q)}(z_\dd)\}$ by increasing grade $q$.
By iterative elimination of the lower rows, corresponding to the fields
with non-zero $U(1)$ charge, one obtains from \flatii\ a system of linear
differential equation of higher order for the first row, $\Pi^\al\equiv\Pi^\al_0$.
These are the well-known Picard-Fuchs equations for the period integrals
$\Pi^\al = \int_{\Ga^\al}\Om$ over the holomorphic $(3,0)$ form on $X$.
Imposing the appropriate boundary conditions, the 
solutions to these equations determine the period vector $\Pi^\al$ from
which all chiral ring coefficients $C_i$ may be obtained
by differentiation. The period vector also determines the flat coordinates
and eventually, one obtains the
holomorphic prepotential $\cx F$ by integration of \cscrii.
This provides an extremely effective means 
to obtain the exact holomorphic data of the topological theory.

Given a topological flat metric on the space $V$ of RR ground states, 
the period vector $\Pi^\al$
determines also the non-holomorphic 
K\"ahler potential $K(z_a,\bb z_{\bb a})$ on $\cx M_{CS}$, and thus the
metric on moduli space \sgCY. The curvature of this metric
is of a restricted form that is compatible with the general 
properties of local $\cx N=2$ supersymmetry in four dimensions \sugra.

\listrefs
\end